\documentclass[prd,
preprintnumbers,amsmath,amssymb]{revtex4}
\usepackage{graphicx}

\DeclareGraphicsExtensions{.pdf,*.jpg}

 \def\be{\begin{equation}}
 \def\ee{\end{equation}}

 \def\r{\rho}
 
 \newcommand{\la}[1]{\label{#1}}
 \newcommand{\ff}{\frac}
 \newcommand{\vev}[1]{{\left< {#1} \right>}}
 \newcommand{\eq}[1]{(\ref{#1})}
 \def\e{\epsilon}
  \def\cN{{\cal N}} \def\cO{{\cal O}}
 \newcommand{\prt}[1]{{\left( {#1} \right)}}

   \def\Lt{{\tilde{L}}}

 \def\t{\tau}

 \def\2{\frac{1}{2}}
 \def\4{\frac{1}{4}}
 \def\Tc{{\theta}}

\catcode`\@=11
\def\@citex[#1]#2{%
\if@filesw \immediate \write \@auxout {\string \citation {#2}}\fi
\@tempcntb\m@ne \let\@h@ld\relax \def\@citea{}%
\@cite{%
  \@for \@citeb:=#2\do {%
    \@ifundefined {b@\@citeb}%
      {\@h@ld\@citea\@tempcntb\m@ne{\bf ?}%
      \@warning {Citation `\@citeb ' on page \thepage \space
undefined}}%
      {\@tempcnta\@tempcntb \advance\@tempcnta\@ne%
      \@tempcntb\number\csname b@\@citeb \endcsname \relax%
      \ifnum\@tempcnta=\@tempcntb 
it
        \ifx\@h@ld\relax%
          \edef \@h@ld{\@citea\csname b@\@citeb\endcsname}%
        \else%
          \edef\@h@ld{\ifmmode{-}\else--\fi\csname
b@\@citeb\endcsname}%
        \fi%
      \else
        \@h@ld\@citea\csname b@\@citeb \endcsname%
        \let\@h@ld\relax%
      \fi}%
    \def\@citea{,\penalty\@highpenalty\,}%
  }\@h@ld
}{#1}}

\def\@citeb#1#2{{[#1]\if@tempswa , #2\fi}}
\def\@citeu#1#2{{$^{#1}$\if@tempswa , #2\fi }}
\def\@citep#1#2{{#1\if@tempswa , #2\fi}}

\begin{document}

\title{Entanglement entropy in a four-dimensional cosmological background}

\author{V. Giantsos 
}
\email{giantsosvangelis@gmail.com}
\author{N. Tetradis 
}
\email{ntetrad@phys.uoa.gr}
\affiliation{ 
Department of Physics,
University of Athens,
University Campus,
Zographou 157 84, Greece}

\begin{abstract}
We compute the holographic entanglement entropy of a thermalized CFT on a
time-dependent background in four dimensions. We consider a slab 
configuration extending beyond the cosmological horizon of 
a Friedmann-Lemaitre-Robertson-Walker metric. We identify a volume term that
corresponds to the thermal entropy of the CFT, as well as terms proportional to
the proper area of the entangling surface which are associated with 
strongly entangled degrees of freedom in the vicinity of this surface or with
the expansion.

\end{abstract}

\maketitle

\section{Introduction}\la{intro}

The entanglement entropy is an interesting observable that quantifies the 
long-range entanglement between classically separated quantum degrees of freedom. 
It is a probe that can characterize the state of matter. 
Calculations of the entanglement entropy rely on the determination of the 
reduced density matrix for the degrees of freedom enclosed by an entangling surface,
by integrating out the degrees of freedom outside the surface.
This density matrix deviates from that of a pure state. 
In this sense, there is an analogy with the thermal entropy, which 
has been exploited in order to study the thermal nature of
field theories in curved spacetimes with horizons.
Explicit calculations of the entanglement entropy 
are possible mainly in low dimensions
\cite{cardy,discretizer,stefan}
and for non-interacting quantum field theories in higher dimensions \cite{sredincki,muller,casini90,pimentel,Kanno:2014lma,Iizuka:2014rua,Kanno:2016qcc}.
However, the holographic approach and the Ryu-Takayanagi proposal
\cite{ryu,review} provide a simpler framework for theories that have a gravitational
dual in the context of the AdS/CFT correspondence \cite{adscft}.
The demanding calculation
of the density matrix is replaced by the simpler calculation of extremizing the area
of a surface that starts from the entangling surface on the boundary of AdS space 
and extends into the bulk.

We are interested in the form and evolution of the entanglement entropy in an 
expanding background. Our motivation comes from the wish to understand its 
possible role in early-time cosmology. For this purpose, we must consider 
time-dependent backgrounds, such as the Friedmann-Lemaitre-Robertson-Walker 
(FLRW) universe.
Also, the de Sitter (dS) spacetime in flat slicing is of interest as the relevant
background for inflation.
The holographic entanglement entropy can be computed in such cases  
through a generalization of the Ryu-Takayanagi proposal for a static background \cite{rtextension}.
Several studies of the entanglement entropy on curved backgrounds 
have been performed, in which  
the connection with the dS gravitational entropy has also been examined \cite{strominger,marolfds,pimentel,fischlerds,Chu:2016uwi,Chugiatag,Ghosh:2018qtg,Giataganas:2019wkd,grieninger,correctionsds,vanderSchee,Park,Manu:2020tty,Giataganas:2020lcj,tetradis}.

Realistic applications to early-time cosmology require the consideration of an
environment with non-zero temperature in the expanding (3+1)-dimensional background.
The analytical calculation of the holographic entanglement entropy in such a setup 
has not been performed up till now. In the following, we generalize the 
zero-temperature results of ref. \cite{tetradis} 
in order to take into account the thermal effects. 
We neglect the spatial curvature of the expanding FLRW background, consistently with
the observed spatial flatness of the Universe. 
In order to introduce a non-zero temperature we consider an asymptotically 
AdS (4+1)-dimensional bulk spacetime that includes a planar black hole.
The dual theory is the $\cN=4 \hspace{0.5em} SU(N)$ sYM theory at a temperature related to
the Hawking temperature. We use appropriate coordinates so that the boundary metric
has the FLRW form. The expansion results in the
redshifting of the physical CFT temperature. 

We consider the entanglement between the
interior of a slab (3-dimensional strip) of comoving width $L$ and its exterior. 
We compute the entanglement entropy through the area of the appropriate 
bulk extremal surface, anchored at the expanding entangling surface on the boundary
\cite{rtextension}. 
Our choice of configuration is made so that explicit analytical expressions can be 
derived for non-zero temperature. However, the result provides all the 
necessary intuition about the dependence of the entanglement entropy on the 
area of the entangling surface, as well as the enclosed volume. It also has an
advantage relative to the spherical entangling surface considered in
ref. \cite{tetradis}. The coordinate transformation that we use in order to 
connect the static problem with the time-dependent one remains valid for 
entangling surfaces that extend beyond the cosmological horizon. This was not
the case for spherical surfaces, for which the transformation was valid up to 
the horizon. 

In the following section we present the calculation of the entanglement entropy for
a slab configuration and zero temperature in order to establish the formalism. This 
section is based on the results of ref. \cite{tetradis}, which are now adapted to 
the slab configuration.
In section \ref{temperature} we generalize the calculation in order to
take into account a thermal environment. The last section includes our conclusions.

\section{AdS$_5$ with FLRW boundary}\la{expansion}

We consider a slicing of (4+1)-dimensional
AdS space that results in a boundary
with a spatially flat (3+1)-dimensional FLRW metric. 
The bulk metric is
\be\la{FRW}
ds^2_5=\ff{1}{z^2}\left[dz^2-N^2(\t,z)d\t^2+A^2(\t,z)d\vec{\bf x}^2\right],
\ee
where
\begin{eqnarray}
N(\t,z) &=& a(\t)\prt{1-\ff{-3\dot{a}^2(\t)+2 a(\t)\ddot{a}(\t)}{4a^4(\t)}\,z^2}
\label{Ntz} \\
A(\t,z) &=& a(\t)\prt{1-\ff{\dot{a}^2(\t)}{4a^4(\t)}\,z^2}.
\label{Atz}
\end{eqnarray}
All quantities are expressed in terms of the AdS length, which
is set equal to 1. The scale factor $a(\t)$ is expressed in terms of conformal time.
The metric is of the typical 
Fefferman-Graham form for an asymptotically AdS space \cite{fg}. 

The bulk metric
can also be written using standard Poincar$\acute{\mbox{e}}$ coordinates as
\be\la{poincare}
ds^2_5=\ff{1}{\zeta^2}\left[d\zeta^2-dt^2+d\vec{\bf x}^2\right].
\ee
The metrics \eq{FRW} and \eq{poincare} are related through the coordinate transformation
\begin{eqnarray}
t(z,\t) &=& \t+\ff{2\dot{a}(\t)a(\t)\,z^2}{-4a^4(\t)+\dot{a}^2(\t)\,z^2}
\label{tzt} \\
\zeta(z,\t) &=& \ff{z}{a(\t)}
\prt{1-\ff{\dot{a}^2(\t)}{4a^4(\t)} \,z^2}^{-1}.
\label{zzt}
\end{eqnarray}

We consider a 3-dimensional slab, with width $L$ in comoving coordinates
 along the $x$-direction and width $\Lt\to \infty$ along the $y$- and $z$-directions. 
 The physical width of the slab grows proportionally to $a(\t)$, 
 following the general expansion.
The space inside the slab is entangled with the exterior. The corresponding 
entanglement entropy at a given time $T$ can be computed through the
extension of the Ryu-Takayanagi proposal \cite{ryu,review}
given in ref. \cite{rtextension}.
The entropy is proportional to the area
\be\la{areadst}
{\rm Area}(\gamma_A)\equiv  I(\e)=\Lt^2\int_\e dx\ff{A^2(\t(x),z(x))}{z^3(x)}\sqrt{A^2(\t(x),z(x))
 -N^2(\t(x),z(x))\prt{\ff{d\t(x)}{dx}}^2+\prt{\ff{dz(x)}{dx}}^2},
\ee
extremized with respect to the functions $\t(x)$ and $z(x)$, with
the boundary conditions $\t(L/2)=\t(-L/2)=T$ and $z(L/2)=z(-L/2)=0$. The integral diverges near the boundary, so that a cutoff must be imposed 
on the bulk coordinate $z$ at $z=\e$.
The entanglement entropy is given by
\be
S=\ff{I(\e)}{4G_5},
\ee
with $G_5$ the bulk Newton's constant.

In order to find the extremum of the area we switch 
to Poincar$\acute{\mbox{e}}$ coordinates, making use
of eqs. \eq{tzt}, \eq{zzt}. The functional becomes
\be\la{areadst}
{\rm Area}(\gamma_A)=\Lt^2\int dx
 \ff{1}{\zeta^{3}(x)}\sqrt{1
 -\prt{\ff{dt(x)}{dx}}^2+\prt{\ff{d\zeta(x)}{dx}}^2}. 
\ee
The solution for the function $t(x)$ is trivial: $t(x)=T={\rm constant}$.
The minimization with respect to $\zeta(x)$ is the standard one for
a slab configuration and a static boundary. The minimal surface is determined by the relation
\be\la{stslab}
\ff{d\zeta(x)}{dx}=\ff{\sqrt{\zeta^6_0-\zeta^6(x)}}{\zeta^3(x)},
\ee
where $\zeta_0$ is a constant that specifies the turning point of the surface, 
that is the deepest point in the bulk that the surface can reach. 
This point corresponds to $x=0$, namely $\zeta(0)=\zeta_0$ 
and $\zeta'(0)=0$. By integrating eq. \eq{stslab} and incorporating the symmetry 
of the surface in the region $-L/2\leq x\leq L/2$, 
we express the slab width in comoving coordinates as
\be\la{width}
L=2\int_0^{\zeta_0}d\zeta \prt{\ff{\zeta}{\zeta_0}}^3\ff{1}{\sqrt{1-\prt{\ff{\zeta}{\zeta_0}}^6}}=\ff{2\sqrt{\pi}\Gamma\prt{\ff{2}{3}}}{\Gamma\prt{\ff{1}{6}}}\zeta_0.
\ee
The extremal surface is given by the implicit relations
\begin{eqnarray}
T &=& \t+\ff{2\dot{a}(\t)a(\t)\,z^2}{-4a^4(\t)+\dot{a}^2(\t)\,z^2}
\label{soltau} \\
\zeta(x) &=& \ff{z}{a(\t)}
\prt{1-\ff{\dot{a}^2(\t)}{4a^4(\t)} \,z^2}^{-1}
\label{solzeta}
\end{eqnarray}
for the functions $\t(x)$ and $z(x)$.

The integration for the area of the extremal surface can be performed by using $\zeta$ as an independent variable. The area becomes \cite{ryu,review}
\be\la{areaeps}
{\rm Area}(\gamma_A)\equiv
 I(\e)=2\Lt^2\ff{1}{\zeta_0^2}\int_{\e_\zeta(T)/\zeta_0}^1dy\ff{1}{y^3\sqrt{1-y^6}},
\ee
where $y=\zeta/\zeta_0$. The integral must be regulated by imposing a 
cutoff on $\zeta$ near the boundary. This cutoff results from the 
cutoff $\epsilon$ imposed on the Fefferman-Graham coordinate $z$.
For $\e\to 0$, we obtain from eq. \eq{soltau}
\be\la{tauex}
\t(\e)=T+\ff{\dot{a}(T)}{2a^3(T)}\e^2 
\ee
near the boundary. By expanding $a(\t)$ in eq. \eq{zzt}, the cutoff on $\zeta$ can be written as
\be\la{cutoff}
\e_\zeta(T)=\ff{\e}{a(T)}\prt{1-\ff{1}{4}H^2(T)\e^2},
\ee
with $H(T)=\dot{a}(T)/a^2(T)$ the Hubble parameter in terms of conformal time.
For $\e\to 0$
the integral of eq. \eq{areaeps} can be evaluated analytically. We find
\be\la{expandfrw}
S=\ff{\Lt^2a^2(T)}{4G_5\e^2}-\ff{\Lt^2L}{8G_5\zeta_0^3}+\ff{\Lt^2a^2(T)H^2(T)}{8G_5}+\cO \prt{\e^2}.
\ee
The time dependence of the scale factor $a(T)$ is arbitrary, as the boundary metric does not result from dynamical equations of motion. However, the above expressions are applicable to physical FLRW cosmologies with metrics that have a dynamical origin. 
Notice that eq. (\ref{width}) implies that 
the second term in eq. (\ref{expandfrw}) is proportional to the ratio $\Lt^2/L^2$ 
and, therefore, independent of $a(T)$.


\section{Temperature}\la{temperature}

In order to include an energy scale in the context of an expanding background
one must abandon the simple framework of a pure AdS bulk. The simplest
generalization involves the presence of a bulk black hole with a mass term $\mu$.
The planar AdS-Schwarzschild black hole in 4+1 dimensions 
is a solution of the Einstein equations with a
negative cosmological constant $\Lambda_5=-6$. 
The metric can be written in Poincar$\acute{\mbox{e}}$ coordinates as
\be\la{eqmetric}
ds^2_5=\ff{1}{\zeta^2}\left[-b(\zeta) dt^2+\ff{d\zeta^2}{b(\zeta)}+d\vec{\bf x}^2\right],\ \ \ \ \ \ b(\zeta)=1-\mu\zeta^4.
\ee
The Hawking temperature of the black hole is $\Tc_0=\mu^{1/4}/\pi=1/(\pi \zeta_h)$, with
$\zeta_h$ the location of the horizon. 
The dual CFT for a static boundary is thermalized at the same temperature.
The above metric can also be expressed as \cite{siopsis,nonis} 
\be\la{FRWm}
ds^2_5=\ff{1}{z^2}\left[dz^2-\tilde{N}^2(\t,z)d\t^2+\tilde{A}^2(\t,z)d\vec{\bf x}^2\right],
\end{equation} 
where
\begin{eqnarray}
\tilde{N}(\t,z) &=& a(\t)\ff{\left[\prt{1-\ff{\dot{a}^2(\t)}{4a^4(\t)}\,z^2}\prt{1-\ff{-3\dot{a}^2(\t)+2a(\t)\ddot{a}(\t)}{4a^4(\t)}\,z^2}\right]-\ff{\mu}{4a^4(\t)}z^4}{\sqrt{\prt{1-\ff{\dot{a}^2(\t)}{4a^4(\t)}\,z^2}^2+\ff{\mu}{4a^4(\t)}\,z^4}}
\hspace{3em} \label{Ntzm} \\
\tilde{A}(\t,z) &=& a(\t)\sqrt{\prt{1-\ff{\dot{a}^2(\t)}{4a^4(\t)}\,z^2}^2+\ff{\mu}{4a^4(\t)}\,z^4}.
\label{Atzm}
\end{eqnarray}
The metrics \eq{eqmetric} and \eq{FRWm} are related through the coordinate transformation
\begin{eqnarray}
t(z,\t) &=& \t-\ff{1}{2\mu^{1/4}}\left[\arctan{\prt{\ff{2\mu^{1/4}a(\t)\dot{a}(\t)z^2}{4a^4(\t)+\prt{2\sqrt{\mu}a^2(\t)-\dot{a}^2(\t)}z^2}}}-{\rm arctanh} \hspace{0.2em} \prt{\ff{2\mu^{1/4}a(\t)\dot{a}(\t)z^2}{-4a^4(\t)+\prt{2\sqrt{\mu}a^2(\t)+\dot{a}^2(\t)}z^2}}\right]\nonumber \\
\label{tztm} \\
\zeta(z,\t) &=& \ff{z}{a(\t)}
\left[\prt{1-\ff{\dot{a}^2(\t)}{4a^4(\t)}\,z^2}^2+\ff{\mu}{4a^4(\t)}\,z^4\right]^{-1/2}.
\label{zztm}
\end{eqnarray}

For the metric \eq{FRWm},
it is instructive to consider the
stress-energy tensor of the dual CFT on the time-dependent boundary, as
determined via holographic renormalization \cite{skenderis}.
We obtain the energy density and pressure
\begin{eqnarray}
\label{eq3te}
\r &=& -\vev{T_{~\t}^{\t}} \hspace{0.5em}=\hspace{0.2em}
\ff{3}{64\pi G_5}\ff{4\mu a^4+\dot{a}^4}{a^8}
\\
P &=& \vev{T_{~x}^{x}} \hspace{0.5em}=\hspace{0.2em}\ff{1}{64\pi G_5}\ff{4\mu a^4+5\dot{a}^4-4a\dot{a}^2\ddot{a}}{a^8}
.
\label{eq3tp}
\end{eqnarray}
The terms proportional to $\mu/a^4$ can be interpreted as the thermal energy density and pressure of a CFT at a temperature $\Tc(T)=\Tc_0/a(T)$, an expression that 
accounts for the 
redshift of the physical temperature by the scale factor.

We turn next to the entanglement entropy of the CFT in the time-dependent setting of eq. \eq{FRWm}.
We consider a 3-dimensional slab of comoving width $L$, as in the zero-temperature case. Similarly to the previous section, the minimization of the area functional can be performed by switching to the coordinates $(t,\zeta)$, in terms of which it reads
\be\la{aream}
{\rm Area}(\gamma_A)=\Lt^2\int dx
\ff{1}{\zeta^{3}(x)}\sqrt{1
-b(\zeta)\prt{\ff{dt(x)}{dx}}^2
+\ff{1}{b(\zeta)}\prt{\ff{d\zeta(x)}{dx}}^2}.
\ee
The minimal curve has the trivial time dependence $t(x)=T={\rm constant}$. 
By setting $z=0$ in eq. \eq{tztm} it can be seen that $T$ corresponds to the value that the function $\t(x)$ on the minimal surface takes when the boundary is reached. 
The integral of motion with respect to $\zeta(x)$ is given by
\be\la{stslabm}
\ff{d\zeta(x)}{dx}=\sqrt{b(\zeta)}\ff{\sqrt{\zeta^6_0-\zeta^6(x)}}{\zeta^3(x)}.
\ee
The comoving slab width is now expressed as
\be\la{widthm}
L=2\int_0^{\zeta_0}d\zeta \prt{\ff{\zeta}{\zeta_0}}^3\ff{1}{\sqrt{b(\zeta)}}\ff{1}{\sqrt{1-\prt{\ff{\zeta}{\zeta_0}}^6}}.
\ee

The extremal surface corresponding to an entangling surface at a time $T$ on the boundary is given by the implicit relations
\begin{eqnarray}
T &=& \t-\ff{1}{2\mu^{1/4}}\left[\arctan{\prt{\ff{2\mu^{1/4}a(\t)\dot{a}(\t)z^2}{4a^4(\t)+\prt{2\sqrt{\mu}a^2(\t)-\dot{a}^2(\t)}z^2}}}-{\rm arctanh} \hspace{0.2em} \prt{\ff{2\mu^{1/4}a(\t)\dot{a}(\t)z^2}{-4a^4(\t)+\prt{2\sqrt{\mu}a^2(\t)+\dot{a}^2(\t)}z^2}}\right]\nonumber \\
\label{soltaum} \\
\zeta(x) &=& \ff{z}{a(\t)}
\left[\prt{1-\ff{\dot{a}^2(\t)}{4a^4(\t)}\,z^2}^2+\ff{\mu}{4a^4(\t)}\,z^4\right]^{-1/2}
\label{solzetam}
\end{eqnarray}
for the functions $\tau(x)$ and $z(x)$.
The integration for the area of the extremal surface can be performed by using $\zeta$ as an independent variable. The area becomes \cite{Erdmenger}
\be\la{areabh}
{\rm Area}(\gamma_A)=I(\e)=2\Lt^2 \int_{\e_{\zeta}(T)}^{\zeta_0}d\zeta
\ff{1}{\zeta^{3}}\ff{1}{\sqrt{b(\zeta)}}\ff{1}{\sqrt{1-
\prt{\ff{\zeta}{\zeta_0}}^6}}.
\ee
The cutoff $\e_{\zeta}$ has been imposed on the bulk coordinate at $\zeta=\e_{\zeta}$.
However, the physical cutoff $\e$ for the time-dependent case
is identified with the one 
imposed on the Fefferman-Graham coordinate $z$ for the metric \eq{FRWm} with
a FLRW boundary. 
The relation between $\e_{\zeta}$ and $\e$ can be determined as in the previous section.
For the asymptotic behavior near the boundary ($\epsilon\to 0$), we obtain from eq. \eq{soltaum}
\be\la{tauexbh}
\t(\e)=T+\ff{\dot{a}(T)}{2a^3(T)}\e^2+\ff{\dot{a}^3(T)}{8a^7(T)}\e^4+\prt{\ff{\mu\dot{a}(T)}{8a^7(T)}+\ff{\dot{a}^5(T)}{32a^{11}(T)}}\e^6+\cO\prt{\e^8}.
\ee
Notice that the black-hole mass induces a correction at the sixth order of the expansion. Such a correction gives vanishing contributions for $\e\to 0$ because the divergences near the horizon are not sufficiently strong in five dimensions. This means that the 
relation between the two cutoffs is still given by eq. \eq{cutoff}.

For a static boundary, 
the width of the strip, the minimal area and the 
resulting entanglement entropy have been calculated in \cite{Erdmenger} through the
evaluation of the integrals (\ref{widthm}), (\ref{areabh}). These 
results can be used in order to obtain the expressions for the case of a FLRW background,
by simply expressing the cutoff $\e_{\zeta}$ in terms of $\e$ through eq. (\ref{cutoff}).
The resulting expressions are given in terms of the Meijer $G$-function:
\begin{eqnarray}
L&=&\ff{\pi\zeta_h}{\sqrt{6}}
G_{5,5}^{3,2}\left(
\begin{array}{c}
\ff{1}{4},\ff{3}{4},\ff{1}{4},\ff{7}{12},\ff{11}{12}\\
\ff{1}{12},\ff{5}{12},\ff{3}{4},0,\ff{1}{2}
\end{array}\middle\vert
\prt{\ff{\zeta_0}{\zeta_h}}^{12}\right)
\label{lexp} \\ 
S&=&\ff{\Lt^2a^2(T)}{4G_5\e^2}
+\ff{\Lt^2L}{4G_5\zeta_0^3} +\ff{\pi\Lt^2\zeta_h}{16\sqrt{6}G_5\zeta_0^3}G_{5,5}^{3,2}\left(
\begin{array}{c}
\ff{3}{4},\ff{5}{4},\ff{1}{4},\ff{7}{12},\ff{11}{12}\\
\ff{1}{12},\ff{5}{12},\ff{3}{4},\ff{1}{2},0
\end{array}\middle\vert
\prt{\ff{\zeta_0}{\zeta_h}}^{12}\right)+\ff{\Lt^2a^2(T)H^2(T)}{8G_5}
+\cO \prt{\e^2}.
\label{eeexp}
\end{eqnarray}
The horizon is related to the Hawking temperature of the black hole
through  $\zeta_h=1/(\pi\Tc_0)$, with $\mu=1/\zeta_h^4$ in five dimensions. 
Notice that the entanglement entropy is proportional to the number of 
degrees of freedom of the theory, as $G_5=\pi/(2N^2)$. 
For vanishing temperature $\Tc_0\to 0$, or $\zeta_h\to \infty$,
this expression reproduces eq. \eq{expandfrw}.

The above expressions become more transparent for large values of the 
comoving slab width $L$, 
corresponding to surfaces that probe the IR and have $\zeta_0 \simeq \zeta_h$. 
The horizon $\zeta_h$, and hence the black-hole temperature $\Tc_0$, are kept fixed. 
In the limit $\zeta_0 / \zeta_h \to 1$ the extremal surface coincides with the
black-hole horizon, apart from the regions near the two endpoints, where it approaches
the boundary. The width $L$ of eq. (\ref{lexp}) 
diverges in this limit. The divergence affects also the area (\ref{eeexp})
of the extremal surface, and appears in the second term in this expression, which is
proportional to $L$. The 
divergence of the first term in the limit $\e \to 0$ is associated with the 
contribution to the area from the vicinity of the boundary. 
The third term in eq. \eq{eeexp} is convergent for $\zeta_0 / \zeta_h \to 1$,
approaching a value that can be found numerically to be $G_h\approx -2.07678$. 
Thus, in the large-width limit the entanglement entropy reads
\be\la{eeexplim}
S=\ff{\Lt^2a^2(T)}{4G_5\e^2}
+\ff{\Lt^2L}{4G_5\zeta_h^3}+G_h\ff{\pi\Lt^2}{16\sqrt{6}G_5\zeta_h^2}
+\ff{\Lt^2a^2(T)H^2(T)}{8G_5}.
\ee
The physical width of the slab and the physical 
scale of the other two dimensions 
grow proportionally to $a(T)$, following the general expansion. 
Hence, the corresponding physical lengths are $a(T)L$ and $a(T)\Lt$. 
Recalling also that the physical temperature of the dual CFT redshifts according to $\Tc(T)=\Tc_0/a(T)$, the above relation can be rewritten as 
\be\la{eelimithawk}
S=\ff{\Lt^2a^2(T)}{4G_5\e^2}+\ff{\pi^3\Lt^2La^3(T)\Tc^3(T)}{4G_5}+G_h\ff{\pi^3\Lt^2a^2(T)\Tc^2(T)}{16\sqrt{6}G_5}+\ff{\tilde{L}^2a^2(T)H^2(T)}{8G_5}.
\ee
From this expression we can read off the entanglement entropy 
for fixed physical lengths $a(T)L$ and $a(T)\Lt$. 
Of course, fixing the physical lengths implies that the 
comoving lengths vary as $1/a(T)$.

The second term in eq. (\ref{eelimithawk}) has the qualitative form of thermal entropy.
It is interesting to compare it with the expected entropy of the thermal CFT.  
The $\cN=4 \hspace{0.5em} SU(N)$ sYM theory contains $n_b=6N^2$ bosons, $n_F=4N^2$ 
Mayorana fermions and $n_V=N^2$ massless vector bosons. Using the standard AdS/CFT 
relation $G_5=\pi/(2N^2)$, the second term in eq. (\ref{eelimithawk}) can be written
as
\be\la{entrsym}
s=\frac{S}{V}=\ff{\pi^3}{4G_5}\Tc^3(T)=\ff{\pi^2}{2}N^2\Tc^3(T)=\ff{3}{4}\prt{1\cdot 6N^2+2\cdot N^2+\ff{7}{8}\cdot 2\cdot 4N^2}\ff{2\pi^2}{45}\Tc^3(T),
\ee
with $V=a^3(T)\Lt^2L$.
The last equality indicates that this term 
can be identified with the thermal entropy.
The factor 3/4 is attributed to the fact that the theory is 
strongly coupled and thus
the expression for the entropy deviates from the perturbative prediction \cite{threefourths}. The first term in eq. (\ref{eelimithawk}) is the 
contribution to the entropy arising from the strong entanglement of 
degrees of freedom in the vicinity of the 
entangling surface. It displays the expected divergence associated with 
short distances on either side of this surface. The entanglement entropy is
proportional to the proper area of the entangling surface, which is now
time-dependent. There is no logarithmic divergence for a slab configuration, so that 
the last two terms are finite contributions proportional
to the area. The first of these terms is temperature-dependent, while the 
last one arises through the expansion \cite{tetradis}. It is noteworthy that  
no volume term associated with purely quantum entanglement arises.

\section{Conclusions}\label{discussion}

Our result for the entanglement entropy of the thermalized CFT for a slab configuration 
in a (3+1)-dimensional expanding background,
given by eqs. \eq{lexp}, \eq{eeexp}, displays several interesting features:

\begin{itemize}
\item
The divergent term has the same form as for a static background, with the
length scale of the entangling surface corresponding to the physical scale
$a(T)\Lt$ that determines the proper area of the surface. This indicates that the divergence is associated with
the entanglement of UV degrees of freedom very close to the entangling surface, 
at distances comparable to the cutoff $\epsilon$.
\item
The effect of the temperature is more transparent for large slab widths, for which
the maximal extension $\zeta_0$ of the extremal surface in the bulk 
coincides with the location of the black-hole horizon $\zeta_h$. This limit can
be viewed also as the high-temperature limit, corresponding to a black hole of
large mass. Eqs. \eq{lexp}, \eq{eeexp} are now reduced to eq. (\ref{eelimithawk}).
The leading thermal contribution to the entropy is proportional
to the volume enclosed by the entangling surface and can be identified with the 
thermal entropy of the CFT. 
There is no volume term associated with purely quantum entanglement.
There is a subleading thermal contribution, proportional to the proper
area of the entangling surface. This contribution is finite, and thus arises
from degrees of freedom beyond the immediate vicinity of the entangling surface. 
However, the entanglement is not strong enough to result in a volume effect.
\item
The finite term associated with the expansion 
depends on the square of the expansion rate and has the same value
both for expanding and contracting backgrounds. It can be interpreted as arising through
degrees of freedom that cross the entangling surface during the expansion or
contraction. However, this process is not strong enough in order to result 
in a volume effect and the term is again proportional to the area of the 
entangling surface. The Hubble parameter $H(T)$ grows very large at the early 
times $T\to 0^+$ of a cosmological expansion driven by matter with an equation of state
$p=w\varepsilon$ with $w>-1/3$.
The finite contribution to the
entanglement entropy within a comoving slab width $L$ scales as
$a^2(T)L^2H^2(T)\sim T^{-2}$.
Physical distances between adjacent points are small in this limit, 
so that the entanglement becomes stronger.
\end{itemize}

As a final comment, we would like to examine the range of applicability of our 
analysis, which is based on the transformations between the coordinates $(t,\zeta)$
for a static boundary and $(\tau,z)$ for a time-dependent one. These transformations
may not be valid over the whole AdS boundary. The discussion of the transformations
(\ref{tztm}), (\ref{zztm}) is too complicated. However, we can obtain the basic
picture by examining the transformations \eq{tzt}, \eq{zzt} for a pure AdS bulk with
a specific boundary.
The choice $a(T)=-(HT)^{-1}$, with constant $H$ and negative conformal time $T$,
corresponds to a de Sitter background. 
The minimal surface is implicitly determined by eqs. \eq{soltau}, \eq{solzeta},
which give
\begin{eqnarray}
T &=& \t\ff{4+H^2z^2}{4-H^2z^2} \label{soltauds} \\
\zeta(x) &=& -T\ff{4Hz}{4+H^2z^2}. 
\label{solzetads}
\end{eqnarray}  
For given values of $L$ and $T$, the maximal value $z_m$ of $z(x)$ on the surface is obtained for $x=0$. Increasing $L$ for fixed $T$ makes
$z_m$ increase, up to the value $z_m=2/H$. As can be seen from eq. (\ref{solzetads}),
this value is attained when $\zeta(0)=\zeta_0=\ff{\Gamma\prt{\ff{1}{6}}}{2\sqrt{\pi}\Gamma\prt{\ff{2}{3}}}L=-T$,
where we have made use of eq. (\ref{width}).
The last relation can be rewritten as 
\be
a(T)L=-\frac{1}{HT}L=\ff{2\sqrt{\pi}\Gamma\prt{\ff{2}{3}}}{\Gamma\prt{\ff{1}{6}}}\ff{1}{H}\simeq 
0.86 \ff{1}{H}.
\label{minm} \ee
This shows that the maximal
physical slab width consistent with the transformations 
is slightly smaller than the horizon radius.
However, the remaining two spatial coordinates do not appear in the transformations
and thus are unconstrained.
The part of the slab that extends beyond the 
cosmological horizon along the corresponding spatial directions
is covered by our analysis. 
Allowing for a non-zero black-hole mass, or a different 
time dependence of the scale factor, does not change
the qualitative conclusions. 
This should be contrasted with the situation for a spherical entangling surface, 
in which the transformations are valid only within the cosmological horizon 
\cite{tetradis}.  In this sense, the current analysis provides an additional step
in our understanding of entanglement entropy in cosmological settings.

\section*{Acknowledgments}
We would like to thank C. Sfetsos for useful discussions.
The research of N. Tetradis was supported by the Hellenic Foundation for
Research and Innovation (H.F.R.I.) under the “First Call for H.F.R.I.
Research Projects to support Faculty members and Researchers and
the procurement of high-cost research equipment grant” (Project
Number: 824).


\end{document}